\documentclass[%
reprint,
 amsmath,amssymb,
 aps,
pra,
]{revtex4-1}

\usepackage{graphicx}% Include figure files
\usepackage{pgfplots}

\usepackage{dcolumn}% Align table columns on decimal point
\usepackage{bm}% bold math
\usepackage{hyperref}% add hypertext capabilities
\usepackage{xcolor}
\usepackage{tikz}
\usetikzlibrary{patterns}
\usepackage{soul}
\usepackage{natbib}
\usepackage{braket}

\newcommand{\qg}{\mathbf{q}}

\newcommand{\kg}{\mathbf{k}}

\begin{document}

%\preprint{APS/123-QED}

\title{Pair formation in quenched unitary Bose gases}% Force line breaks with \\
\author{S. Musolino}
 \email{s.musolino@tue.nl}
\affiliation{Eindhoven University of Technology, P.O. Box 513, 5600 MB Eindhoven, The Netherlands}
\author{V. E. Colussi}
\affiliation{Eindhoven University of Technology, P.O. Box 513, 5600 MB Eindhoven, The Netherlands}

\author{S. J. J. M. F. Kokkelmans}
\affiliation{Eindhoven University of Technology, P.O. Box 513, 5600 MB Eindhoven, The Netherlands}

\date{\today}% It is always \today, today,

\begin{abstract}
We study a degenerate Bose gas quenched to unitarity by solving a many-body model including three-body losses and correlations up to second order.  As the gas evolves in this strongly interacting regime, the buildup of correlations leads to the formation of extended pairs bound purely by many-body effects, analogous to the phenomenon of Cooper pairing in the BCS regime of the Fermi gas.  Through fast sweeps away from unitarity, we detail how the correlation growth and formation of bound pairs emerge in the fraction of unbound atoms remaining after the sweep, finding quantitative agreement with experiment.  We comment on the possible role of higher-order effects in explaining the deviation of our theoretical results from experiment for slower sweeps and longer times spent in the unitary regime.      
\end{abstract}

\pacs{Valid PACS appear here}% PACS, the Physics and Astronomy
\maketitle

%\tableofcontents
%\linenumbers
\section{\label{sec:level1}Introduction}
%Intro
In ultracold quantum gases, precision control of magnetically tunable Feshbach resonances makes it possible to tune the effective interaction strength, characterized by the $s$-wave scattering length $a$~\cite{art:chin}.  As $a$ becomes much larger than the interparticle spacing $n^{-1/3}$, where $n$ is the atomic density, the gas enters the unitary regime $(n|a|^3\gg1)$. At unitarity ($|a|\to\infty$), interactions between atoms are as strong as allowed by quantum mechanics. Moreover, the macroscopic properties of unitary quantum gases appear insensitive to microscopic physics and therefore paradigmatic for other strongly correlated systems, including the inner crust of neutron stars and the quark-gluon plasma~\cite{0034-4885-72-12-126001,PhysRevLett.91.102002}.   The universality of the unitary Fermi gas has been both theoretically and experimentally well established over the past two decades~\cite{book:Zwerger}.  Under the universality hypothesis, the unitary Bose gas is also expected to behave similarly, with thermodynamic properties and relations that scale continuously solely with the \textquotedblleft Fermi\textquotedblright scales constructed from powers of $n$, including the Fermi wave number $k_n=(6\pi^2 n)^{1/3}$, energy $E_n=\hbar^2 k_n^2/2m$, and time $t_n=\hbar/E_n$, where $m$ is the atomic mass~\cite{art:ho}. 

%Efimov and experimental summary
Unlike their fermionic counterparts, at unitarity three bosons may form an infinite series of bound Efimov trimers~\cite{art:braaten2006} with characteristic finite size set by the three-body parameter $\kappa_*$ ~\cite{art:3bp,art:3bp2,PhysRevA.99.012702}.  Whereas Pauli repulsion suppresses three-body losses for fermions, the Efimov effect leads to a catastrophic $a^4$ scaling of three-body losses near unitarity, and therefore the unitary Bose gas is inherently unstable.  In Refs.~\cite{art:makotyn,art:klauss, art:eigen, art:eigenpretherm}, this barrier was overcome through a fast quench from the weakly interacting to the unitary regime, where the establishment of a steady state was observed before heating dominates.  Time-resolved studies of the single-particle momentum distribution in Ref.~\cite{art:eigenpretherm} revealed that the theoretically predicted prethermal state~\cite{PhysRevA.93.033653,art:sykes,PhysRevA.90.063626} transitions to steady state prior to being overcome by heating.  Although these findings, combined with studies of loss dynamics in Refs.~\cite{art:makotyn,art:klauss, art:eigen}, are consistent with the universality hypothesis, a macroscopic population of Efimov trimers was observed in Ref.~\cite{art:klauss}. Understanding the role of the Efimov effect~\cite{art:colussi_three,art:efimovquench} and dynamics of higher-order correlations~\cite{art:kira,art:colussimk,PhysRevA.98.053612} in the quenched unitary Bose gas remains, however, an ongoing pursuit in the community.

%ramp discussion and connection with BEC-BCS
The difficulties of probing the system at unitarity require that experiments return to the more stable and better-understood weakly interacting regime. During the course of the experiment, we have to distinguish different types of atomic pairs: (i) pairs of atoms with opposite momentum, analogous to Cooper pairs in Fermi gases, (ii) {\it embedded dimers}  at unitarity whose size is determined by the mean interparticle separation, and (iii) weakly bound molecules away from unitarity, whose size is determined by the scattering length.

\begin{figure}
\centering
\includegraphics[width=8.6cm]{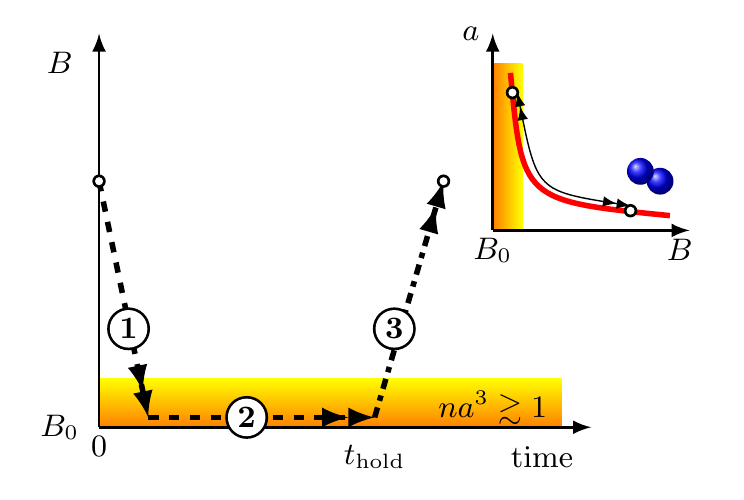}
\caption{Schematic representation of the experimental protocol used 
in Refs.~\cite{art:makotyn,art:klauss, art:eigen, art:eigenpretherm}. First, the magnetic field  $B$ is ramped suddenly towards the resonant value $B_0$, taking the system from the weakly interacting ($n a^3< 1$) to the unitary regime $n a^3 \gtrsim 1$ (shaded region).  In the second stage, the system evolves at unitarity for a variable time $t_\mathrm{hold}$. In the third and final step, the system is ramped back from unitarity with
a different ramp rate (proportional to the slope of line 3)
away from resonance and returns back to the weakly interacting regime where measurements are made and weakly bound molecules can be found. Inset: Feshbach resonance with the scattering length
$a$ as a function of $B$ during the sequence represented in the
main figure.}
\label{fig:scattB}
   \end{figure}

According to the experimental procedure of Refs.~\cite{art:makotyn,art:klauss, art:eigen, art:eigenpretherm}, illustrated in Fig.~\ref{fig:scattB}, a Bose gas is initially quenched from the weakly interacting to the unitary regime, held there for a variable time $t_\mathrm{hold}$, and finally probed again in the weakly interacting regime.  Here, the size of a molecule is much smaller than the mean interparticle separation, and the distinction between unbound and bound atoms is physically meaningful again~\cite{art:kohler_adia}. In the unitary regime, unbound pairs progressively localize onto the scale of the interparticle spacing, purely due to many-body effects~\cite{art:colussimk}. 
 The nature of these embedded dimers is reflected by a universal time-dependent size $a_\mathrm{eff}$, fit to the universal form
\begin{equation}
k_n a_\mathrm{eff}=1.58 +3.44 \left(\frac{t_n}{t_\mathrm{hold}}\right)^2,
 \label{eq:aeff}
\end{equation}
which indicates a transition from unbound ($a_\mathrm{eff}\to\infty$) to bound ($a_\mathrm{eff}\sim k_n^{-1}$) on Fermi timescales, as we will discuss in Sec.~\ref{subsec:embed}.
It is interesting to note the analogy of pair formation in the quenched unitary Bose gas to pair formation in the unitary Fermi gas~\cite{art:stoof2009}, which is at the center of the so-called BCS-BEC crossover.
When entering this crossover from the Bardeen-Cooper-Schrieffer (BCS) side, fermionic pairs, loosely bound by the medium, smoothly evolve into tightly bound molecules that are stable even without the medium, when passing through to the Bose-Einstein condensation (BEC) side, while the effective atomic interaction changes from attractive to repulsive~\cite{book:Zwerger}.
For these experiments, a very successful technique was employed utilizing fast magnetic field sweeps to effectively project the fermionic pairs onto molecules throughout the whole crossover regime~\cite{art:regal_cond, art:hodby, art:altman, art:matyja,rew:Ketterle,art:reviewCrossover}.

%Describe our approach
In this work, we quench an initially pure Bose condensate to unitarity and track the resultant dynamics up to the level of two-body correlations, while including universal three-body losses phenomenologically.  We then model the final step shown in Fig.~\ref{fig:scattB} by a fast-sweep projection technique in the spirit of Ref.~\cite{art:altman}, count the number of remaining unbound atoms, and compare quantitatively our results with the experimental findings of Ref.~\cite{art:eigen}. Unlike in the experiment, in our model, we are able to distinguish between three-body losses and formation of molecules when determining the number of remaining unbound atoms.  Through this ability, we estimate the universal three-body loss-rate coefficient by refitting the experimental data of Ref.~\cite{art:eigen}. 
We also compare the predictions of our model for the number of unbound atoms with the results of that work, finding generally good agreement for fast ramp rates and for slower ramp rates at earlier times ($t_\mathrm{hold}\lesssim 0.5 t_n$).  As correlations grow and the condensate becomes increasingly depleted for longer times spent in the unitary regime, we highlight the dominant contribution of the embedded dimers in the number of unbound atoms detected after fast-sweep projection away from unitarity. 

The organization of this work is as follows.  In Sec.~\ref{sec:model}, we outline our many-body model (Sec.~\ref{subsec:manybodyeq}), adapt the technique of fast-sweep projection from Ref.~\cite{art:altman} (Sec.~\ref{subsec:fastsweep}) for Bose gases, and develop the theory of bound pairs in the unitary regime discussed in Ref.~\cite{art:colussimk} (Sec.~\ref{subsec:embed}).  In Sec.~\ref{sec:3body}, three-body losses are introduced phenomenologically into our many-body model, and in Sec.~\ref{sec:comparison}, we discuss the results of our model and compare them with the experimental findings of Ref.~\cite{art:eigen}.  We conclude in Sec.~\ref{sec:conclusion} and comment on prospects for future study.

\section{\label{sec:model} Model}
\subsection{\label{subsec:manybodyeq} Many-body equations}
We model a uniform gas of identical
spinless bosons interacting via pairwise interactions described by the single-channel many-body Hamiltonian
\begin{equation}
\hat{H}=\sum_\kg \frac{\hbar^2 k^2}{2m}\hat{a}_\kg^\dagger\hat{a}_\kg+
\sum_{\kg, \kg', \qg} V_{\kg, \kg', \qg} \hat{a}_{\kg+\qg}^\dagger\hat{a}_{\kg'-\qg}^\dagger
\hat{a}_{\kg'}\hat{a}_{\kg},
 \label{eq:hamiltonian}
\end{equation}
where  $V_{\kg, \kg', \qg}= (g/2)\zeta(\kg-\kg'+2\qg)
\zeta^\ast(\kg-\kg')$ is a non local separable potential with interaction strength $g$, step-function form factor $\zeta(\kg)=\theta (\Lambda-
|\kg|/2)$, and finite cutoff $\Lambda$, giving rise to a finite-range interaction both in momentum and position space.
This model is suitable for describing open-channel dominated Feshbach resonances, which includes all degenerate unitary Bose gas experiments to date~\cite{art:makotyn,art:fletcher, art:klauss,art:eigen, art:eigenpretherm}.  To fix the free parameters of the separable potential, we first set the strength of the potential, $g=U_0 \Gamma$,  where $U_0=4\pi \hbar^2 a/m$ and $\Gamma =(1- 2a\Lambda/\pi)^{-1}$, to reproduce the exact two-body $T$ matrix in the zero-energy limit~\cite{art:colussimk, art:kokk}.  To fix $\Lambda$, we follow Ref.~\cite{art:colussimk} and set $\Lambda=2/\pi \bar{a}$ to obtain finite-range corrections to the binding energy of the Feshbach molecule $E_\mathrm{b}\simeq
-\hbar^2/m(a-\bar{a})^2$, valid only to first order in $1/\Lambda a$, and where $\bar{a}=0.955 r_\mathrm{vdW}$ is the mean scattering
length that depends on the van der Waals length $r_\mathrm{vdW}$, for a particular atomic species~\cite{art:chin}.  Consequently, at unitarity we obtain a finite interaction strength $g=-\pi^3\hbar^2\bar{a}/m$ for $a\to\infty$.

To model the condensate and excitations, we make the Bogoliubov approximation~\cite{art:bosereview} and decompose the operator $\hat{a}_\kg=\psi_\kg +\delta \hat{a}_\kg$ with  $\braket{\delta \hat{a}_\kg}=0$.  We assume that only the atomic condensate is macroscopically occupied so that $\braket{\hat{a}_\kg}=\psi_0 \delta_{\kg,0}$ and consider only fluctuations of the excitations.  These assumptions are valid provided the excited modes are not macroscopically occupied. Furthermore, we build our many-body theory from the cumulant expansion~\cite{art:kira, art:colussimk}, which separates clusters of correlated particles within an interacting many-body system. Here, we consider only up to second-order clusters (correlations), described by the condensate wave function $\psi_0$ and the one-body $\rho_\kg\equiv \braket{\hat{a}_\kg^\dagger
\hat{a}_\kg}$ and pairing $\kappa_\kg\equiv \braket{\hat{a}_{-\kg}\hat{a}_\kg}$ density matrices for excitations~\cite{book:blaizot}.

We derive the Hartree-Fock Bogoliubov (HFB) equations~\cite{book:blaizot} from the Heisenberg equation of motion for one or two $\hat{a}$-operator products and evaluate the expectation values in cumulant expansion, neglecting clusters of three or more particles. Summarily,  if $\hat{\mathcal{O}}$ is a specific operator, using  $i\hbar \braket{d\hat{\mathcal{O}}/dt}=\braket{ [\hat{\mathcal{O}}, \hat{H}]}$ one has
\begin{eqnarray}
 i\hbar \dot{\psi}_0&=& g \left(|\zeta(0)|^2|\psi_0|^2+ 2\sum_{\kg \neq 0} |\zeta(\kg)|^2\rho_\kg\right) \psi_0 \notag\\ 
 &+& g \psi_0^\ast \sum_{\kg\neq 0}\zeta(0)\zeta^\ast(2\kg) \kappa_\kg,
 \label{eq:psia3}\\
  \hbar \dot{\rho}_\kg &=& 2\mbox{Im}\left[\Delta_\kg\kappa_\kg^\ast\right],
  \label{eq:gn3}\\
    i\hbar\dot{\kappa}_\kg &=& 2h_\kg\kappa_\kg+ \left(1 + 2\rho_\kg\right)\Delta_\kg,\label{eq:kappa}
\end{eqnarray}
where
\begin{eqnarray}
h_\kg= \frac{\hbar^2 k^2}{2m} + 2g \left(|\zeta(\kg)|^2|\psi_0|^2 +\sum_{\qg\neq 0}|\zeta(\kg -\qg)|^2\rho_\qg\right)
\end{eqnarray}
and 
\begin{eqnarray}
 \Delta_\kg=g\zeta(2\kg)\left(\zeta^\ast(0)\psi_0^2 +\sum_{\qg\neq 0}\zeta^\ast(2\qg) \kappa_\qg\right)
\end{eqnarray}
are the Hartree-Fock Hamiltonian and the pairing field, respectively~\cite{book:blaizot}. The HFB theory results in a mean-field description, typically suitable for the weakly interacting regime where $n|a|^3\ll1$. However, here we formulate a finite-range HFB theory, which yields a finite mean-field energy $g$ at unitarity, as discussed above. We argue that this theory is applicable for strong interactions since $g$ remains small respect to $E_n$, or equivalently $nr_\mathrm{vdW}^3\ll1$. Note that this condition is well satisfied for all experiments in the unitary regime to date ($nr_\mathrm{vdW}^3<10^{-5}$)~\cite{PhysRevLett.87.120406,art:colussimk,art:makotyn,art:klauss, art:eigen, art:eigenpretherm}.

To simulate the first two steps of the experimental sequence illustrated in Fig.~\ref{fig:scattB},
Eqs.~\eqref{eq:psia3}-\eqref{eq:kappa} are solved at fixed initial density $n_\mathrm{in}= N_\mathrm{in}/V$, where $N_\mathrm{in}\equiv N(t=0)$ with total atom number $N(t)$ in a volume $V$~\footnote{The Fermi scales $k_n$, $E_n$, and $t_n$ are defined in terms of the initial density $n_\mathrm{in}$.}. In particular, we consider experiments done in a box-trap, modelled as a uniform system~\cite{art:gaunt}. We begin at $t=0$ from a pure condensate with $|\psi_0|^2= n_\mathrm{in}$.  The scattering length is then ramped over $2 $ $\mu$s to unitarity, where the system evolves for a varying amount of time, $t_\mathrm{hold}$.  As the gas evolves at unitarity and in the absence of losses, the condensate fraction becomes depleted as correlated pair excitations are generated and counted by $\rho_\kg$ as studied in Ref.~\cite{art:sykes}. We expect that the increase of $\rho_\kg$ beyond unity makes higher-order cumulants strongly driven and their inclusion in the model cannot be justified. 
Therefore, following Ref.~\cite{art:colussimk}, we  restrict our analysis to $t\leq2 t_n$ where $\rho_\kg < 1$ remains valid.
%fastsweep
\subsection{\label{subsec:fastsweep} Fast-sweep projection away from unitarity}
 We finally model the third step of Fig.~\ref{fig:scattB} with a projection of the many-body state at unitarity onto a molecular state at finite scattering length and count the number of molecules.  Intuitively, in the limiting case of a sudden switch of the magnetic field, the number of molecules may be calculated, to good approximation, by simply projecting the state at unitarity onto molecules at the final magnetic field $B_\mathrm{end}$. For finite ramp rates $R=-dB/dt$, this approximation is not valid.  In this case, the number of molecules may be calculated approximately by projection onto an effective molecular state $\phi_\ast$ with scattering length $a_\ast$ larger than the final scattering length $ a_\mathrm{end}$ and intermediate to both the sudden and adiabatic cases,
  as detailed in Ref.~\cite{art:altman}. This method provides an indirect measure of the buildup of correlations at unitarity. The conceptual problem of bound pairs in the unitary regime is revisited in Sec.~\ref{subsec:embed}.

We construct a compound bosonic operator
 \begin{equation}
\hat{b}_0^\dagger \equiv \sum_{\kg} \frac{\phi_\ast(k)}{\sqrt{2}} \hat{a}^\dagger_{-\kg} \hat{a}^\dagger_{\kg},
 \label{eq:b0dag}
\end{equation}
counting molecules away from unitarity with zero center of mass and relative momentum $\kg$ of the constituent
atoms,  where $\phi_\ast(k)$ is a molecular wave function with a finite scattering length $a_\ast$ whose value will be specified shortly. By construction, the $\hat{b}$ operator satisfies $[\hat{b}_0, \hat{b}_0]=[\hat{b}_0^\dagger, \hat{b}_0^\dagger]= 0$, and the canonical commutation relation $[\hat{b}_0, \hat{b}_0^\dagger]= 1 + \sum_\kg |\phi_\ast(k)|^2 (\hat{a}^\dagger_{\kg}\hat{a}_{\kg} + 
 \hat{a}^\dagger_{-\kg}\hat{a}_{-\kg})$ is approximately well satisfied $\braket{[\hat{b}_0, \hat{b}_0^\dagger]}\simeq 1$ away from unitarity, where the molecules are spatially much smaller than the interparticle spacing. We note also that the approach of counting composite bosons [Eq.~\eqref{eq:b0dag}] has been also used extensively for counting fermionic pairs along the BEC-BCS crossover~\cite{art:nozieres, art:altman, art:peralifast}.

%This seems more like a possible footnote:  where $a$-operators satisfy fermionic %anticommutation and the $\sqrt{2}$-factor in Eq.~\eqref{eq:b0dag} is omitted because %of the spin degree of freedom.

We evaluate the expectation value $\braket{\hat{b}^\dagger_0 \hat{b}_0}$, which can be expanded in terms of first- and second-order cumulants, consistently with the theory presented in Sec.~\ref{subsec:manybodyeq}. Therefore, the molecular fraction is
 \begin{equation}
 \begin{split}
   \frac{2N_\mathrm{mol}}{N_\mathrm{in}}&=V\sum_\kg |\phi_*(k)|^2\left(|\psi_0|^4\delta_{\kg, 0} + \frac{2}{V}\rho_\kg^2 \right)\\
   &+V\left| \sum_{\kg} \phi_*(k) \kappa_\kg^\ast\right|^2\\
   &+V\sum_\kg\,2\mbox{Re}\left[\phi_*(0) \,[\psi_0^{\dagger}]^2\, \kappa_\kg\, [\phi_*(k)]^\ast\right],
   \end{split}
 \label{eq:n0_n}
\end{equation}
where $N_\mathrm{in}/2$ is the total possible number of molecules.  At $t_\mathrm{hold}=0$ immediately following the completion of the quench, $|\psi_0|^2\approx n$ and $\rho_\kg\approx\kappa_\kg\approx 0$, and therefore only the first term on the right-hand side of Eq.~\eqref{eq:n0_n} contributes.  This contribution can be interpreted as the overlap of the molecular wave function and the atomic mean field~\cite{art:kohler_adia} and scales as $n a_{\ast}^3$ proportional to the ratio of atomic and molecular volumes.   This overlap must be insignificant so that $n a_{\ast}^3<1$, and molecules can be separated from the many-body background.  The remaining terms in Eq.~\eqref{eq:n0_n} measure the overlap between molecular and pairing wave functions~\cite{art:pairwave} and reflect the development of correlations as the gas evolves in the unitary regime.  We note that Eq.~\eqref{eq:n0_n} is in agreement with the first-quantized multichannel description in position space found in Ref.~\cite{art:kohler_adia}.

%Seems like more of a footnote:  Respect to the equivalent version for %fermions~\cite{art:altman}, here, we have extra terms dependent on the atomic %condensate. 

In the evaluation of Eq.~\eqref{eq:n0_n}, the molecular wave function has the universal form 
\begin{equation}
 \phi_\ast(k) =\frac{\sqrt{\mathcal{N}a_\ast^{3}}  }{1+(k a_\ast)^2},
 \label{eq:psistar}
\end{equation}
valid provided $a_\ast\gg r_\mathrm{vdW}$ \cite{art:chin}.  The normalization constant $\mathcal{N}=4 \pi^2/\{\arctan(\Lambda a_\ast) - \Lambda a_\ast/[1+(\Lambda a_\ast)^2]\}$ ensures that $\sum_\kg^\Lambda 
|\phi_\ast(k)|^2=1$. As mentioned above, Eq.~\eqref{eq:psistar} is written in terms of $a_\ast$, whose value depends on $R$. $a_\ast$ represents the point at which the evolution of the system under the ramp changes from sudden to adiabatic, and the creation and dissociation of molecules is halted~\cite{art:altman,art:matyja}.   Quantitatively, this occurs when $E_\mathrm{b}/\hbar=E_\mathrm{b}^{-1}\dot{E_\mathrm{b}}$ is satisfied, where $E_\mathrm{b}= -\hbar^2/m(a-\bar{a})^2$ is the molecular binding energy including finite-range effects \cite{art:chin}.  We obtain specific values of $a_\ast$  from the real solution of the third-order polynomial equation
\begin{equation}
\left(a_\ast - a_\mathrm{bg}\right)^2 \left(a_\ast - \bar{a}\right)
 =\frac{\hbar \Delta B a_\mathrm{bg}}{2mR},
 \label{eq:astar}
\end{equation}
where $a_\mathrm{bg}$ is the background scattering length, and $\Delta B$ is the width of the Feshbach resonance~\cite{art:chin}\footnote{To derive Eq.~\eqref{eq:astar}, the two-channel expression $a=a_\mathrm{bg}- m G^2/(8\pi \hbar^2 \nu)$ was used, where $G$ is the coupling between the two channels, $\nu=\Delta \mu (B-B_0)$ is the detuning between collision energy and bound-state energy in the closed channel, and $\Delta\mu$ is the difference in magnetic moments between the two channels~\cite{rew:duinestoof, art:servaas_ram}. This introduces an explicit dependence on $1/R$.}.
In the sudden limit, the initial state is projected onto the final scattering length $a_\mathrm{end}\equiv a(B_\mathrm{end})\gg a_\ast$ which is on the order of $a_\mathrm{bg}$ or $\bar{a}$ for $1/R\to 0$.

\begin{figure}
\centering
 \includegraphics[width=8.6cm]{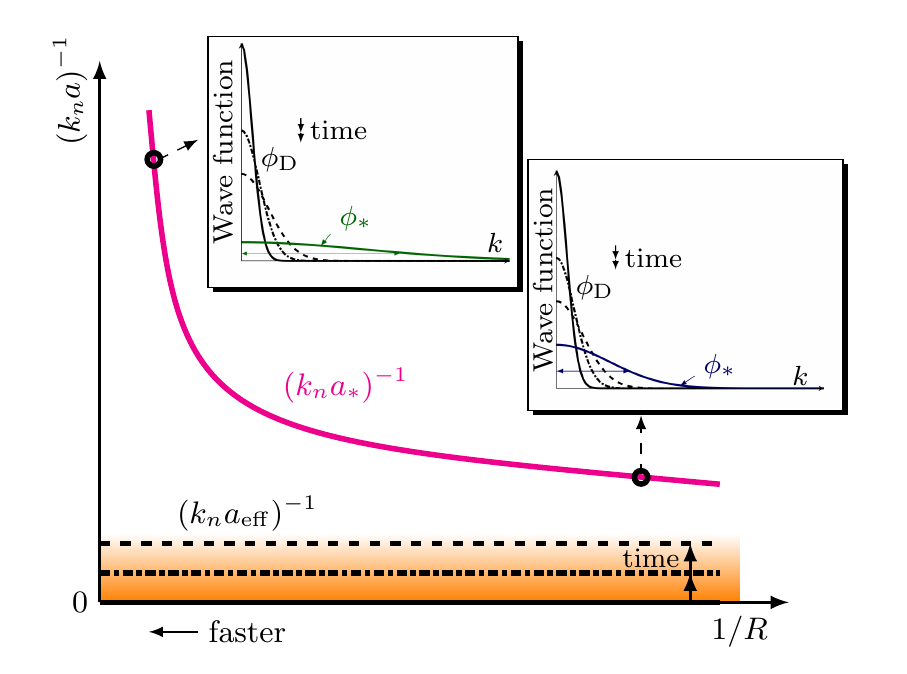}
  \caption{A qualitative illustration of the variation of $a_\ast$ (solid red lines) with the ramp rate $1/R$ and of the variation of $a_\mathrm{eff}$ (solid and dashed black lines) with increasing $t_\mathrm{hold}$ as indicated by arrows in the shaded region.  The direction of faster ramps is indicated explicitly.  In the insets, the molecular $[\phi_\ast(k)]$ (solid green and blue lines) and bound pair $[\phi_\mathrm{D}(k)]$ (solid, dash-dotted, and dashed lines) wave functions are compared for increasing $t_\mathrm{hold}$ indicated by arrows.}
\label{fig:over}
    \end{figure}

The dependence of $a_\ast$ on the ramp rate is shown by the solid red line in Fig.~\ref{fig:over}.  Generally, larger values of $a_\ast^{-1}$ indicate a faster ramp and the many-body state at unitarity is projected onto more localized molecules.  Consequently, $\phi_\ast(k)$ will be less pronounced at low momenta than for slower ramps, which can be seen in the insets of Fig.~\ref{fig:over}.
%embedded
\subsection{\label{subsec:embed} Embedded dimers at unitarity}
To link the buildup of correlations at unitarity with the fast-sweep production of molecules, it is instructive to introduce a many-body length scale that can be compared with $a_\ast$. Here, we follow the approach outlined in Ref.~\cite{art:colussimk} and study embedded two-body bound states at unitarity. 

To obtain the spectrum of these dimers embedded in the unitary Bose gas, the homogeneous part of Eq.~\eqref{eq:kappa}, including only terms dependent on $\kappa_\kg$, is solved as a two-body Schr\"odinger equation in the quasi-stationary limit~\cite{art:colussimk,art:kira}.  This approach is valid provided $\kappa_\kg$ evolves faster than the density dynamics and scattering among clusters (inhomogeneous terms) are  neglected, in which case one obtains an eigenvalue equation
\begin{eqnarray}
E_\mathrm{2B}^{(\nu)} \phi_\nu^\mathrm{R} (\kg)&=& 2 h_\kg\phi_\nu^\mathrm{R} (\kg) + (1+2\rho_\kg)\sum_{\qg\neq 0}g\zeta(2\kg)\notag\\ &\times& \zeta^\ast(2\qg) \phi_\nu^\mathrm{R} (\qg),  
 \label{eq:eqmotion}
\end{eqnarray}
where $E_{2B}^{(\nu)}$ is a two-body eigenenergy and  $\phi_\nu^\mathrm{R}(\kg)$ is a right-handed wave function ~\cite{art:kira, art:colussimk}. 
%The index $\nu$ indicates 
The left-handed wave function $\phi_\nu^\mathrm{L}(\kg)$ is related via $\phi_\nu^\mathrm{R}(\kg)=(1+2\rho_\kg)\phi_\nu^\mathrm{L}(\kg)$, and they satisfy the usual orthogonality $\sum_\kg [\phi_\nu^\mathrm{L} (\kg)]^\ast 
\phi_\mu^\mathrm{R} (\kg)=\delta_{\nu, \mu}$ and normalization $\sum_\nu [\phi_\nu^\mathrm{L} (\kg)]^\ast \phi_\nu^\mathrm{R} (\qg)=\delta_{\kg, \qg}$ conditions.

It is illustrative to compare Eq.~\eqref{eq:eqmotion} with the Schr\"odinger equation for a Cooper pair in the BEC-BCS crossover, which depends instead on the Pauli-blocking factor $(1-2\rho_\kg)$ ~\cite{art:nozieres, art:reviewCrossover}. Whereas the blocking factor in the BEC-BCS crossover theory forbids scattering at occupied intermediate states~\cite{book:fetter}, the intermediate states for a Bose gas are Bose enhanced~\cite{PhysRevA.57.1230}.  Both effects may lead to weakly bound pairs which are held together purely by many-body effects, whose presence was predicted in the finite-temperature phase diagram of the strongly interacting Bose gas~\cite{art:stoof2009}. 

Following Ref.~\cite{art:colussimk}, we track the gradual development in time of these embedded dimers, solutions of Eq.~\eqref{eq:eqmotion} with wave function $\phi_\mathrm{D}(k)$ and binding energy $E_{2B}^{\mathrm{D}} \equiv -\hbar^2/m a_\mathrm{eff}^2$. Equation~\eqref{eq:eqmotion} can be solved numerically, yielding  $E_\mathrm{2B}^D$ as a function of time. The evolution of $E_{2B}^{\mathrm{D}}$ was fit in Ref.~\cite{art:colussimk}, and we quote that result in Eq.~\eqref{eq:aeff}.
 Initially, these dimers are basically unbound ($a_\mathrm{eff}\sim\infty$), but through the subsequent buildup of correlations and quantum depletion they are localized ($a_\mathrm{eff}\propto k_n^{-1}$) onto the Fermi scale and behave universally.

Comparing $a_\ast$ with $a_\mathrm{eff}$ provides a convenient way of characterizing the underlying physics of the fast-sweep projection.  These scales are shown in Fig.~\ref{fig:over}, where the development of $a_\mathrm{eff}^{-1}$ as the gas evolves in the unitary regime is represented by the progression of horizontal lines in the shaded region.  As discussed in Sec.~\ref{subsec:fastsweep}, the fast-sweep projection must be such that $k_n a_\ast \ll 1$ and therefore outside of the shaded region.  These length scales may also be used to understand how the buildup of correlations influences the number of remaining unbound atoms after the fast-sweep projection.  The evolution of $\phi_\mathrm{D}(k)$ with $t_\mathrm{hold}$ is shown along with $\phi_\ast(k)$ in Fig.~\ref{fig:over} for two different ramp rates.  The gradual localization of $\phi_\mathrm{D}(k)$ onto the Fermi scale leads to increasing overlap with $\phi_\ast(k)$.  This behavior is more pronounced for slower ramps and for longer $t_\mathrm{hold}$.  Therefore, we intuitively expect that embedded dimers make an increasing contribution to the overlap term in Eq.~\eqref{eq:n0_n} and therefore the number of molecules produced by the fast-sweep projection.

\begin{figure}
  \centering
 \includegraphics[width=8.6cm]{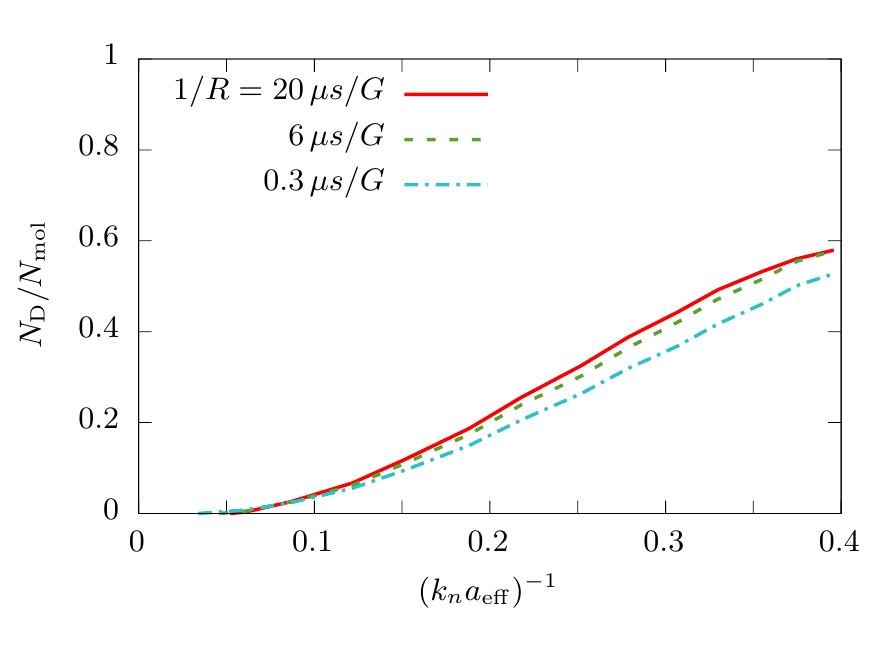}
\caption{Contribution of the embedded dimers formed at unitarity to the total number of molecules produced by the fast-sweep projection away from the unitary regime shown for three ramp rates within the range of experimental interest.  Time is implicit in the inverse effective scattering length in the sense of Eq.~\eqref{eq:aeff}.  By $t_\mathrm{hold}\sim 2t_n$ when $(k_n a_\mathrm{eff})^{-1}\simeq0.4$, we obtain a maximum contribution $N_\mathrm{D}/N_\mathrm{mol}\approx 0.6$.  }
\label{fig:ndnmol_aeff}
\end{figure}

To determine the role of the embedded dimers at unitarity, we decompose $\kappa_\kg$ in the basis of $\phi_\nu^\mathrm{R}(\kg) $ as
\begin{equation}
 \kappa_\kg = \sum_\nu c_\nu \phi_\nu^\mathrm{R}(\kg) \Leftrightarrow c_\nu = 
  \sum_\kg [\phi_\nu^\mathrm{L}(\kg)]^\ast \kappa_\kg,
  \label{eq:kappa_spec}
\end{equation}
where the coefficient $c_\nu$ quantifies the relative weight of the component $\nu$
within the total $\kappa_\kg$. 
We define the embedded dimer contribution $N_\mathrm{D}$ in Eq.~\eqref{eq:n0_n}, by evaluating only the component $\nu=D$ of Eq.~\eqref{eq:kappa_spec}
\begin{equation}
\begin{split}
\frac{2 N_\mathrm{D}}{N_\mathrm{in}}&= V \left|\sum_\kg \phi_\ast(k) [\phi_\mathrm{D}^\mathrm{R}(\kg)]^\ast
\sum_\qg [\kappa_\qg]^\ast \phi_\mathrm{D}^\mathrm{L}(\qg)\right|^2\\ 
&+ V\sum_\kg\,2\mbox{Re}\Big[\phi_*(0)[ \psi_0^\dagger]^2
\left(\sum_\qg [\phi_\mathrm{D}^\mathrm{L}(\qg)]^\ast \kappa_\qg \right)\\
&\times \phi_\mathrm{D}^\mathrm{R}(\kg)[\phi_*(k)]^\ast \Big].
\end{split}
 \label{eq:nemb_pro}
\end{equation}
Figure~\ref{fig:ndnmol_aeff} shows the ratio between embedded dimers and total number of molecules as a function of $(k_n a_\mathrm{eff})^{-1}$ after 
fast-sweep projections for three ramp rates of experimental interest.  We find that by $t_\mathrm{hold}\sim 2 t_n$, when $(k_n a_\mathrm{eff})^{-1}\sim 0.4$, embedded dimers make up $\approx60\%$ of the detected molecules. Therefore, the fast-sweep projection increasingly converts embedded dimers into weakly bound molecules away from resonance, as the gas spends more time at unitarity, agreeing with the intuitive overlap picture shown in Fig.~\ref{fig:over}.  We note that the behavior shown in Fig.~\ref{fig:ndnmol_aeff} is reminiscent of the monotonic conversion of fermions pairs into molecules along the BEC-BCS crossover as a function of the scattering length~\cite{art:salasnich, art:reviewCrossover}.
%as unitarity is approached from negative scattering lengths.   
%In general, $F_\mathrm{D}$ increases in time and at
%$t\simeq 2 t_n$ it is around $60\%$, consistently with~\cite{art:kira}.
%threebodylosses
%threebodylosses
\section{\label{sec:3body}Modeling three-body losses}
The development of strong correlations at unitarity is also accompanied by strong losses~\cite{art:makotyn,art:fletcher, art:klauss,art:eigen, art:eigenpretherm,art:efimovquench,art:sykes}.  In Ref.~\cite{art:eigenpretherm}, by focusing on the early-time dynamics of the tail of the single-particle momentum distribution for $k/k_n\gtrsim 0.8$,  it was possible to experimentally distinguish between the formation of a steady-state and long-time heating. However, this separation was not possible experimentally in Refs.~\cite{art:makotyn,art:eigen} for observables depending on the full range of momentum.  In the present work, we model the findings of Ref.~\cite{art:eigen} and study the number of unbound atoms detected following  the completion of the experimental sequence illustrated in Fig.~\ref{fig:scattB}.  In particular, the number of unbound atoms decreases in time because of two main phenomena, which are difficult to distinguish experimentally: molecular formation and three-body losses.  Therefore, the inclusion of losses is required to make a quantitative comparison.

We assume a universal form for three-body losses scaling as $n^{2/3}$ in degenerate Bose gases, ignoring possible log-periodicities due to the Efimov effect~\cite{art:efimovquench,art:eismann}
\begin{align}
 \frac{\dot{N}(t)}{N(t)} &=-\frac{A}{t_{n}}\label{eq:ntot_unit}.
 \end{align}
 This gives an {\it effective} three-body loss coefficient 
  \begin{align}
K_3^\mathrm{eff}(n(t))&=\frac{A \hbar(6 \pi^2)^{2/3}}{2 m}n(t)^{-4/3},\label{eq:L3N}
\end{align}
 that satisfies the standard relation $\dot{N}(t)/N(t)=-K_3^\mathrm{eff}N(t)^2/V^2$ for a uniform system~\cite{art:chin}.  We treat the constant $A$ as a free parameter that is varied in Sec.~\ref{sec:comparison} in order to fit experimental data of Ref.~\cite{art:eigen}.  The form of Eq.~\eqref{eq:ntot_unit} was found experimentally in Refs.~\cite{art:eigen,art:klauss} and is theoretically motivated by the universal substitution $a^4\to a_\mathrm{eff}^4$ in the scaling law of $K_3^\mathrm{eff}$ for shallow dimers as was suggested in Refs.~\cite{art:colussi_three,art:Stoof2014}.
 For clarity, the universal scaling $n^{2/3}$ was found to be valid only for $t_\mathrm{hold} \lesssim 4 t_n$, for longer time the loss rate scales as $n^{26/9}$, consistent with the results for a thermal gas at unitarity~\cite{art:petrov_werner, PhysRevLett.110.163202} and beyond the limit of validity of our model, as discussed in Sec.~\ref{subsec:manybodyeq}.

To incorporate three-body losses into the HFB equations, we consider the time derivative of the   atomic density $dn(t)/dt=d(N(t)/V)/dt=d(|\psi_0(t)|^2)/dt + \sum_{\kg \neq 0} d\rho_\kg(t)/dt$ and in order to satisfy Eq.~\eqref{eq:ntot_unit} we only have to modify Eqs.~\eqref{eq:psia3} and~\eqref{eq:gn3} with additional terms 
 \begin{eqnarray}
 i\hbar \dot{\psi}_0&=& \cdots -i\dfrac{\hbar}{2}K^\mathrm{eff}_3(n(t))\, n^2(t)\psi_0,
  \label{eq:psia3_l3}\\
 \hbar\dot{\rho}_\kg&=&\cdots -\hbar K_3^\mathrm{eff}(n(t))\, n^2(t) \rho_\kg,
  \label{eq:gn3_l3}
 \end{eqnarray}
 where $\cdots$ represents the lossless terms of the HFB equations.  We note that a similar phenomenological approach has been used at the level of the Gross-Pitaevksii equation in Ref.~\cite{art:Ancilotto2015} and also to describe the Bosenova in Refs.~\cite{art:altin, art:snyder}.  These approaches, however, did not include density dependence in the three-body loss coefficient.  We also note that it should be possible to go beyond this phenomenological approach through a proper inclusion of third-order correlations into an extension of the many-body model outlined in Sec.~\ref{sec:model}.  These matters remain the subject of future study.

%numerics
\section{ \label{sec:comparison}Results}
In this section, we compare the results of our model to the experimental data in Ref.~\cite{art:eigen}.  We approximate the box cylindrical trap used in that work as a homogeneous gas~\cite{art:gaunt} and numerically solve the HFB equations including losses [Eqs.~\eqref{eq:kappa},~\eqref{eq:psia3_l3}, and ~\eqref{eq:gn3_l3}] for the $^{39}$K Feshbach resonance at $B=402$ G with $a_\mathrm{bg}=-29 a_0$, $\Delta B=-52 $ G, and $\bar{a}=61.7 a_0$~\cite{art:chin}.  To mimic the experimental setup, we fix the initial density $n_\mathrm{in}$ and simulate up to $t_\mathrm{hold}=2t_n$, which is the range of validity of our model as discussed in Sec.~\ref{sec:model}.  We then calculate the total number of unbound atoms after the fast sweep away from unitarity from Eq.~\eqref{eq:n0_n} for the ramp rates used experimentally.  We calculate the number of free (unbound) atoms as 
 \begin{equation}
  N_\mathrm{free}(t_\mathrm{hold}, 1/R) = N(t_\mathrm{hold}) -2 N_\mathrm{mol}(t_\mathrm{hold}, 1/R),
  \label{eq:natoms}
 \end{equation}
where the ramp-rate dependence is indicated explicitly.  

%validity considerations
Before discussing the results, we comment on the validity of our approach.  For the $^{39}$K Feshbach resonance at $B=402$ G, we find that for the ramp rates and initial densities considered $2.0  \leq (k_n a_\ast)^{-1} \leq 6.7$, and therefore the fast-sweep projection method outlined in Sec.~\ref{subsec:fastsweep} can be applied.  Although not analyzed in this work, we estimate that this method can also be applied to model the fast-sweep projection studied in Ref.~\cite{art:klauss} with $^{85}$Rb~\footnote{For the $^{85}$Rb Feshbach resonance at $B=155 $ G studied in Ref.~\cite{art:klauss} with $a_\mathrm{bg}=-443 a_0$, $\Delta B=10.7 $ G, and $\bar{a}=79.1 a_0$~\cite{art:chin}, we find that for initial densities between $0.2$ and $5.8 \times 10^{12}\, \mbox{cm}^{-3}$ the range is $1.6 \leq (k_n a_\ast)^{-1}\leq 4.7$ for $1/R=12.5 \mu s/G$.}.  For smaller $1/R$ and hence smaller $a_\ast$, we follow in the spirit Ref.~\cite{art:matyja} and check the expression of $E_\mathrm{b}$ used to calculate Eq.~\eqref{eq:astar} against a coupled-channel calculation~\cite{thomas}, finding discrepancies of less than $5\%$.

 \begin{figure}
  \centering
 \includegraphics[width=8.6cm]{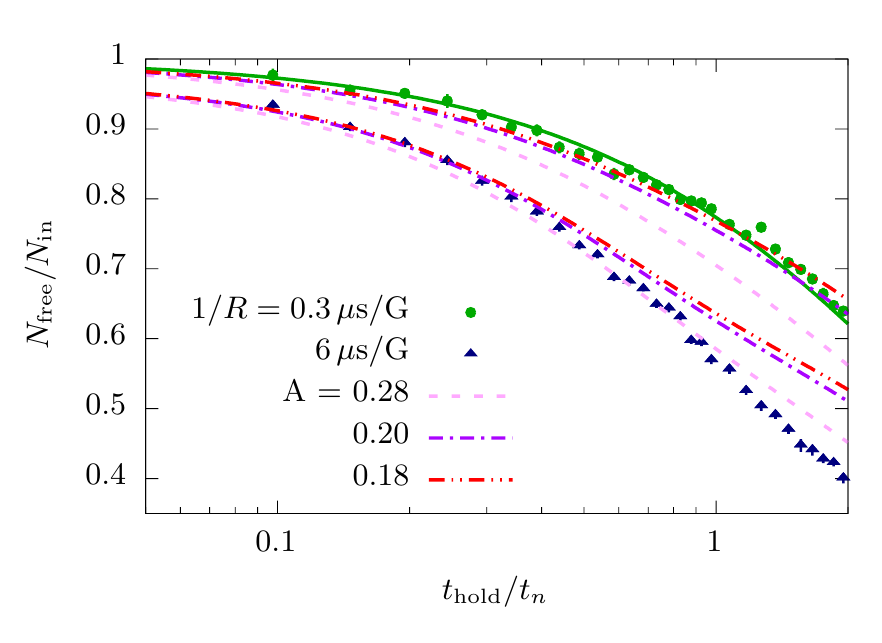}
\caption{Fraction of unbound atoms remaining after fast-sweep projection away from unitarity as a function of $t_\mathrm{hold}/t_n$ for $n_\mathrm{in}=2.7\times 10^{12}$ cm$^{-3}$, where $t_n=41$ $\mu$s.  The experimental data points are taken from Ref.~\cite{art:eigen}.  Assuming that the $0.3$ $\mu$s/G ramp projects the gas at unitarity only onto unbound atoms and taking $A=0.28$ yields the solid green line.  The different colored theoretical curves correspond to $A=\{0.28,0.20,0.18\}$ (pink dashed, purple dot-dashed, and red dot-dot-dashed lines, respectively).}
\label{fig:nobs_thold}
\end{figure}

\begin{figure}
  \centering
 \includegraphics[width=8.6cm]{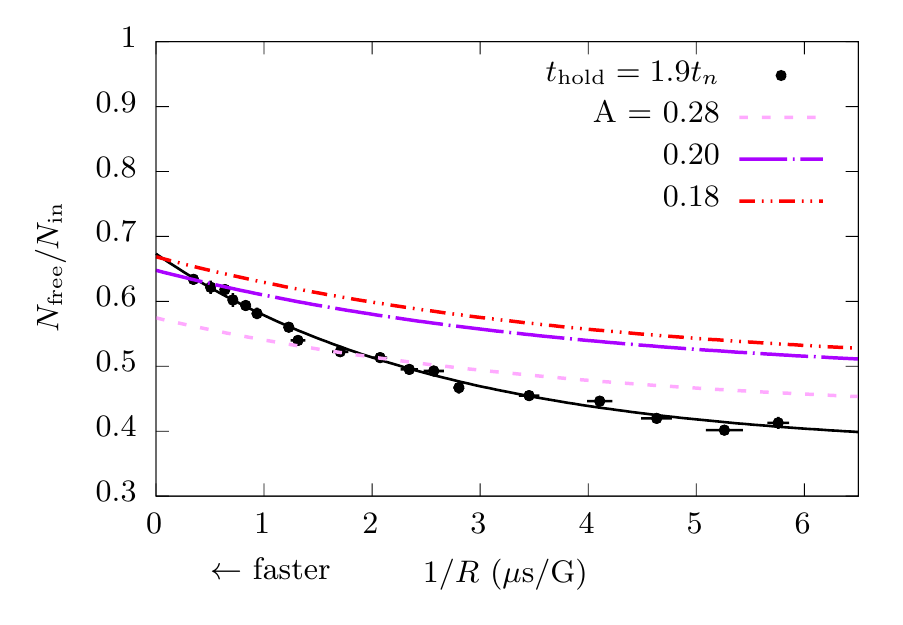}
\caption{Fraction of unbound atoms produced after a fast-sweep projection away from unitarity over a range of ramp rates and fixed $t_\mathrm{hold}=80$ $\mu$s $\approx1.9t_n$ and initial density $n_\mathrm{in}=2.7\times 10^{12}$ cm$^{-3}$.  Here, we compare theoretical results for $A=\{0.28,0.20,0.18\}$ (pink dashed, purple dot-dashed, and red dot-dot-dashed lines, respectively) as indicated in the legend.  The experimental results from Ref.~\cite{art:eigen} are indicated by the data points along with the Landau-Zener exponential fit with $\gamma^{-1}= 2.2 $ $\mu$s/G (black solid line) as discussed in the main text.}
\label{fig:nobs_rate}
\end{figure}

%discussion of Fig. 4
Our results for $N_\mathrm{free}$ are compared against the experimental findings of Ref.~\cite{art:eigen} as a function of $t_\mathrm{hold}$ for initial density $n_\mathrm{in}=2.7\times 10^{12}$ cm$^{-3}$ and ramp rates $0.3$ and $6$ $\mu$s/G, as shown in Fig.~\ref{fig:nobs_thold}.  At $t_\mathrm{hold}=0$, the small gap between the theoretical results for the two different ramp rates is due solely to the first term on the right-hand side of Eq.~\eqref{eq:n0_n} which scales as $n_\mathrm{in} a_{\ast}^3$ and therefore varies with the ramp rate [see Sec.~\ref{subsec:fastsweep}].  At later times, pair correlations begin to develop, and the overlap between embedded dimers at unitarity $[\phi_\mathrm{D}(k)]$ with molecules away from resonance $[\phi_\ast(k)]$ increases, as illustrated in Fig.~\ref{fig:over}.  Consequently, the decrease of $N_\mathrm{free}$ shown in Fig.~\ref{fig:nobs_rate} is due jointly to molecular formation and three-body losses.  

The constant $A$ was estimated in Ref.~\cite{art:eigen} for $^{39}$K as $A=0.28(3)$ by assuming that the $0.3$ $\mu$s/G ramp projects the gas at unitarity only onto unbound atoms [solid green line in Fig.~\ref{fig:nobs_thold}].  In our model, we separate the contributions of molecular formation and loss, and it is therefore possible to test this assumption and provide an independent estimation of $A$ using the approach outlined in Sec.~\ref{sec:3body}.  We therefore adjust $A$ in the HFB equations including losses [Eqs.~\eqref{eq:kappa},~\eqref{eq:psia3_l3}, and ~\eqref{eq:gn3_l3}], and {\it refit} the $0.3$ $\mu$s/G experimental data as shown in Fig.~\ref{fig:nobs_thold}.  For this specific ramp, we find a molecular fraction $\approx10\%$, which is compatible with the experimental estimate in Ref.~\cite{art:eigen}.  By comparing three values $A=\{0.28,0.20,0.18\}$ to the $0.3$ $\mu$s/G experimental data, we find that $A=0.20$ provides the best fit of the experimental results over the full range of $t_\mathrm{hold}$ considered in this work.  For the slower $6 $ $\mu$s/G ramp, we find that $A=0.20$ gives good agreement at early times until roughly $t_\mathrm{hold}\gtrsim 0.5 t_n$. We discuss possible sources of this discrepancy at longer $t_\mathrm{hold}$ at the conclusion of this section.

    \begin{figure}
      \begin{minipage}{.5\textwidth}
\centering
 \includegraphics[width=8.6cm]{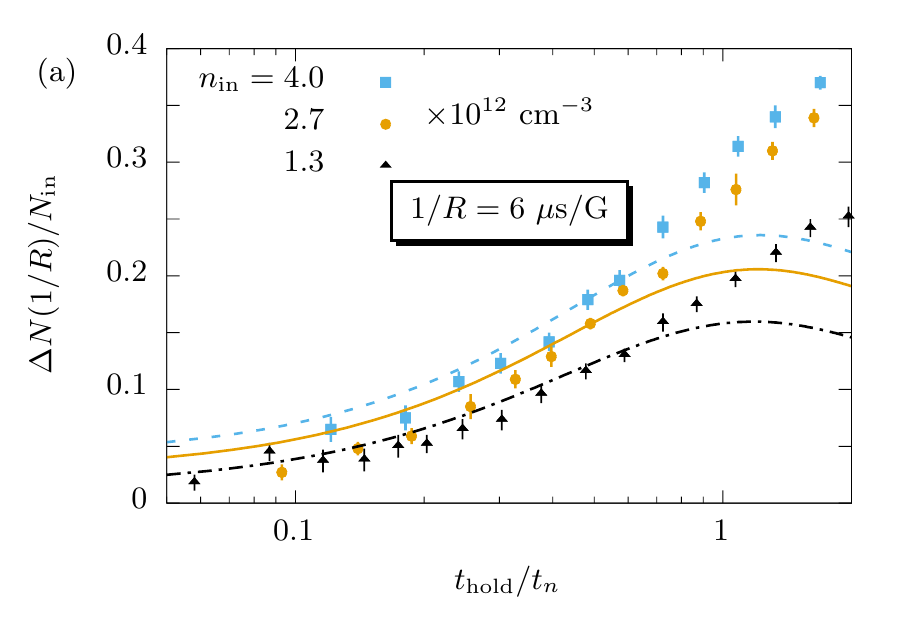}
      \end{minipage}
      \begin{minipage}{.5\textwidth}
        \centering
 \includegraphics[width=8.6cm]{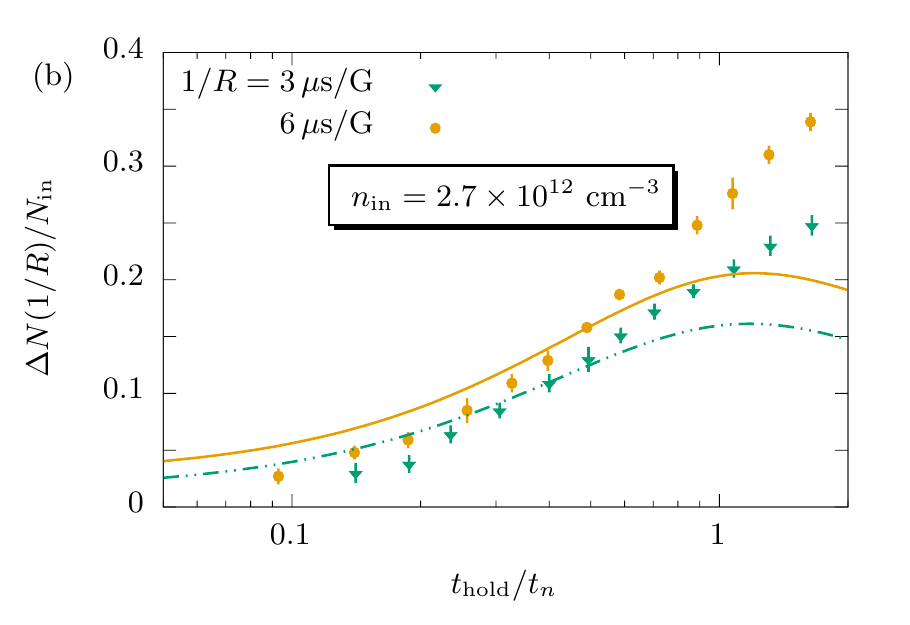}
       \caption{Difference in the fraction of unbound atoms for three different densities and for two different ramp rates measured by $\Delta N (1/R)$ over a range of $t_\mathrm{hold}$.  Our theoretical predictions (lines) are compared against the experimental results (data points) from Ref.~\cite{art:eigen}.  (a) Behavior of $\Delta N (6\, \mu \mathrm{s/G})$  over a range initial densities $n_\mathrm{in}=4.0$, $2.7$, and $1.3\times 10^{12} $ cm$^{-3}$ [$t_n=32,$ $41,$ and $66$ $\mu$s, respectively] as indicated by color (blue dashed, yellow solid, and black dot-dashed lines, respectively).  (b) Behavior of $\Delta N (3\, \mu \mathrm{s/G})$ and $\Delta N (6\, \mu \mathrm{s/G})$ for fixed initial density $n_\mathrm{in}=2.7\times 10^{12} $ cm$^{-3}$ as indicated by color (green dot-dot-dashed and yellow solid lines, respectively).}\label{fig:dn_thold_n}
      \end{minipage}
    \end{figure}
%discussion of Fig. 5 
Our results for $N_\mathrm{free}$ over a range of $1/R$ are compared against the experimental findings in Ref.~\cite{art:eigen} as shown in Fig.~\ref{fig:nobs_rate}.  The results shown in Fig.~\ref{fig:nobs_rate} are at fixed $t_\mathrm{hold}=1.9 t_n$, nearing the limit of validity of our model [see Sec.~\ref{subsec:manybodyeq}]. The intuitive picture, discussed in Secs.~\ref{subsec:fastsweep}, \ref{subsec:embed} and illustrated in Fig.~\ref{fig:over} provides a way to understand our results particularly at this later time where the bound pairs at unitarity play a dominant role (see Fig.~\ref{fig:ndnmol_aeff}).  For smaller ramp rates, the largest values of $N_\mathrm{free}$ shown in Fig.~\ref{fig:nobs_rate} result from the fast-sweep projection occurring further away from unitarity where the overlap between embedded dimers $[\phi_\mathrm{D}(k)]$ with molecules $[\phi_\ast(k)]$ becomes minimal.  We find good agreement with experiment only for the fastest ramps considered using the refitted value $A=0.20$.  In Ref.~\cite{art:eigen} the ramp-rate dependence of $N_\mathrm{free}$ is fit to a Landau-Zener exponential $N_\mathrm{free}= \alpha+ \beta\ \mbox{exp}(-\gamma/R)$~\cite{art:hodby}, where they found $1/\gamma=2.2(3) $ $\mu$s/G.  From fitting the $A=0.20$ theoretical data in Fig.~\ref{fig:nobs_rate} (dot-dashed purple curve), we find $1/\gamma= 4.1 $ $\mu$s/G. The possible sources of discrepancy for slower ramps will be discussed at the end of this section.

%Begin discussion of Fig. 6
In addition, we analyze $N_\mathrm{free}$ over a range of initial densities, $n_\mathrm{in}$, and compare against the experimental results in Ref.~\cite{art:eigen}. Taking the refitted value $A=0.20$, we follow experiment and vary $n_\mathrm{in}$ between $1.3\times 10^{12}$ and $4.0\times 10^{12} $ cm$^{-3}$, measuring the difference $\Delta N(1/R)\equiv N_\mathrm{free}(t_\mathrm{hold},0.3\,\mu \mathrm{s/G})-N_\mathrm{free}(t_\mathrm{hold}, 1/R)$ as shown in Figs.~\ref{fig:dn_thold_n}(a), and~\ref{fig:dn_thold_n}(b).  We note that for fixed $A$ this is equivalent in our model to the difference $\Delta N(1/R)=2(N_\mathrm{mol}(t_\mathrm{hold}, 1/R)-N_\mathrm{mol}(t_\mathrm{hold},0.3\,\mu \mathrm{s/G}))$.  At $t_\mathrm{hold}=0$, $\Delta N$ is nonzero due to the first term of Eq.~\eqref{eq:n0_n} scaling as $n_\mathrm{in} a_{\ast}^3$ that was also discussed earlier in connection with Fig.~\ref{fig:nobs_thold}.  At later times, the gradual separation of the $\Delta N$ curves shown in Fig.~\ref{fig:dn_thold_n}(a) can be understood by comparing the density-dependent and independent length scales $a_\mathrm{eff}$ and $a_\ast$, respectively.  The many-body length scale $a_\mathrm{eff}\propto n_\mathrm{in}^{-1/3}$ is sensitive to changes in the initial density, whereas $a_\ast$ remains fixed by the ramp rate $1/R$.  Consequently, the overlap between $\phi_\mathrm{D}(k)$ and $\phi_\ast(k)$ increases with $n_\mathrm{in}$, which results in the separation of the theoretical $\Delta N$ curves in Fig.~\ref{fig:dn_thold_n}(a), where $1/R=6\, \mu$s/G.  In Fig.~\ref{fig:dn_thold_n}(b), we also compare our results for $\Delta N$ at fixed $n_\mathrm{in}$ for ramp rates $3$ and $6 $ $\mu$s/G, in order to differentiate between $1/R$ and $n_\mathrm{in}$ dependencies.  As before, we attribute the separation of the theoretical $\Delta N$ curves to the time dependence of the overlap between $\phi_\mathrm{D}(k)$ and $\phi_\ast(k)$ and the dominance of the bound pairs at unitarity at later times [see Fig.~\ref{fig:ndnmol_aeff}]. This separation is reflected also in the experimental data shown in Figs.~\ref{fig:dn_thold_n}(a) and~\ref{fig:dn_thold_n}(b).  In general, our predictions in Figs.~\ref{fig:dn_thold_n}(a) and (b) match the experimental data well until we begin to underestimate $\Delta N$ compared to experiment at times $t_\mathrm{hold} \gtrsim 0.5 t_n$.

%
%how do we understand the LZ disagreement?  Fast probes large momentum, small lengthscales.  Slow probes opposite...
%beginning of our interpretation of the discrepancy at later times for slower ramps.  
We now address the deviation between our theoretical predictions presented in this section and the experimental results of Ref.~\cite{art:eigen} for the $3$ and $6$ $\mu$s/G ramps over longer timescales $t_\mathrm{hold}\gtrsim 0.5 t_n$.  In Ref.~\cite{art:eigen}, it was experimentally observed that a degenerate Bose gas quenched to the unitary regime undergoes a universal crossover to the thermal regime by $t_\mathrm{hold}/t_n\approx 4.0$.  In the thermal regime, the three-body loss rate $\dot{N}/N$ scales as $n^{26/9}$~\cite{PhysRevLett.110.163202}.  However, the quantitative agreement between theory and experiment for the loss-dominated $0.3$ $\mu$s/G ramp shown in Fig.~\ref{fig:nobs_thold} is consistent with the $2/3$ power law in the degenerate regime [see Eq.~\eqref{eq:ntot_unit}]. 

In Ref.~\cite{art:eigenpretherm}, it was experimentally observed that momentum modes with $k/k_n\gtrsim 0.8$ reach a prethermal steady state and plateau by $t_\mathrm{hold}\sim t_n$ before long-time heating dominates. In our model the momentum modes described by $\rho_\kg$ in the HFB equations (see Sec. II A) do not plateau as function of $t_\mathrm{hold}$ but oscillate in time, as in Refs.~\cite{gao2018universal,heras2018early,art:sykes}, where the dynamics at unitarity is described through a time-dependent coherent-state pairing wave function ansatz equivalent to the HFB model~\cite{PhysRevA.90.021602, PhysRevA.91.013616}.  This would be most apparent for the slowest $6$ $\mu$s/G ramp [see Fig.~\ref{fig:dn_thold_n}(a)] where $2.5\leq(k_n a_\ast)^{-1}\leq3.2$, and therefore it is possible that the physics behind the plateau are responsible for the deviation between theory and experiment.

Finally, from the experimental findings in Ref.~\cite{art:klauss}, a macroscopic population of Efimov trimers, corresponding to $8\%$ of the initial state, was found after performing a fast-sweep projection away from unitarity.  To estimate the potential relevance of Efimov trimers, we follow Refs.~\cite{art:colussi_three,art:efimovquench} and compare the Fermi scale with the size of the nearby first-excited trimer $R_{3b}^{(1)}=(1+s_0^2)^{1/2}e^{\pi/s_0}/(3/2)^{1/2}\kappa_*$, where $s_0\approx 1.00624$ and $\kappa_*=0.226/r_\mathrm{vdW}$ is the universal three-body parameter \cite{art:braaten2006,art:3bp,art:3bp2}.  For the density range considered in Fig.~\ref{fig:dn_thold_n}, we estimate that $1.7\leq k_n R^{(1)}_\mathrm{3b}\leq 2.5$.  Based on the qualitative findings in Ref.~\cite{art:efimovquench},  the first-excited Efimov trimer
population is expected to grow more slowly than the dimer contribution to the molecular fraction, and this may be partially responsible for the deviation at later times~\footnote{Here, we reference specifically Figs. S3(b) and S3(c) in Ref.~\cite{art:efimovquench}.}.  However, in that work a breakdown of the Landau-Zener behavior was found for increasing $t_\mathrm{hold}$, which qualitatively disagrees with the experimental and theoretical results shown in Fig.~\ref{fig:nobs_rate} displaying this behavior.  We leave, however, the possibility of resolving this deviation by either including into our many-body model three-body correlations or equilibrating collisions \cite{PhysRevA.98.053612} as inspiration for future work. 
%
%
%
%Conclusion
\section{\label{sec:conclusion}Conclusion}
In this work, we present a dynamical model of the degenerate Bose gas quenched to unitarity, which we compare against recent experimental results \cite{art:eigen} for the number of unbound atoms remaining after a fast-sweep ramp away from the unitary regime.  We adopt the method of Ref.~\cite{art:altman} from the study of Cooper pairs in the BEC-BCS crossover and project the many-body state in the unitary regime onto molecular states away from unitarity.  As the Bose gas evolves in the unitary regime, the buildup of correlations and quantum depletion leads to the formation of pairs bound purely by many-body effects as studied in Ref.~\cite{art:colussimk}.  The size of these embedded dimers sets a new length scale given by the effective scattering length, and we draw the  analogy with Cooper pairing in BCS theory \cite{art:reviewCrossover}.  We find that this length scale and the development of the bound pairs at unitarity provide an intuitive way to frame both the theoretical results of our model and the experimental results of Ref.~\cite{art:eigen} for the number of unbound atoms remaining after a fast-sweep projection.  In order to make a quantitative comparison with the experiment, we include three-body losses phenomenologically in our many-body model by assuming an effective universal three-body loss-rate coefficient and by refitting the experimental estimate of this parameter.  

We find good quantitative agreement with experimental data from Ref.~\cite{art:eigen} for the fastest ramp considered in that work over the full range of times where our model remains valid.  However, for slower ramps we begin to deviate quantitatively from the experimental findings at later times.  We argue that this deviation may be due to the presence of Efimov trimers or from the equilibrating effect of collisions both of which are not described in our model.  This motivates further development of our theoretical model to include higher-order correlations, which remains a subject of ongoing study.
\section{\label{sec:acknowledgments}Acknowledgments}
The authors thank Christoph Eigen, Zoran Hadzibabic, and Robert P. Smith for inspiring discussion and for providing experimental data. We also acknowledge Thomas Secker, Paul Mestrom, and Denise Braun for useful discussion.
This work is supported by Netherlands Organization for Scientific Research (NWO) under Grant No. 680-47-623.

\bibliographystyle{apsrev4-1}
\bibliography{biblio_paper}
\end{document}